\documentclass{article}
\def\p{\partial}
\def\s{\sigma}

\def\g{\gamma}

\def\d{\delta}
\def\de{\delta}
\def\D{\Delta}
\def\ld{\lambda}
\def\Ld{\Lambda}
\def\L{\Lambda}
\def\ep{\epsilon}
\def\e{\eta}

\def\rh{\rho}
\def\z{\zeta}

\def\b{\beta}

\def\a{\alpha}

\def\pdellx'{\frac{\partial}{\partial x'}}
\def\pdellw'{\frac{\partial}{\partial w'}}

\newcommand{\be}{\begin{equation}}
\newcommand{\ee}{\end{equation}}
\def\bed{\begin{displaymath}}
\def\eed{\end{displaymath}}
\def\bea{\begin{eqnarray}}
\def\eea{\end{eqncrray}}
\def\[{$$}
\def\]{$$}

\begin{document}
\title{Quantum Yang-Mills gravity in flat space-time and 
effective curved space-time for motions of classical objects}

\author{JONG-PING HSU\\
Physics Department,
 University of Massachusetts Dartmouth \\
 North Dartmouth, MA 02747-2300, USA.}


\maketitle
{\small  Yang-Mills gravity with translational gauge group T(4) in flat 
space-time implies a simple self-coupling of gravitons and
a truly conserved energy-momentum tensor.  Its consistency with 
experiments crucially depends on an interesting property that an `effective Riemannian 
metric tensor' emerges in and only in the geometric-optics limit of the photon and particle wave 
equations.  We obtain Feynman rules for a coupled graviton-fermion system, 
including a general graviton propagator with two gauge parameters and
the interaction of ghost particles.  The equation of motion of 
macroscopic objects, as an N-body system, is demonstrated as the geometric-optics limit of the fermion 
wave equation.  We discuss a relativistic Hamilton-Jacobi equation 
with an `effective Riemann metric tensor' for the classical particles.}
\bigskip


 
\section{ Introduction}

Classical Yang-Mills gravity with 
translation gauge symmetry in flat space-time was shown to be
consistent with known 
experiments, including classical tests and the gravitational 
quadrupole radiation.\cite{1,2,3}  The key for this interesting consistence with 
experiments (see Appendix) is that an `effective Riemann metric tensor' 
emerges in the classical limit of particles' wave equation.    That is, the 
Hamilton-Jacobi equation of motion turns  out to involve an 
effective metric tensor $G_{\mu\nu}$ 
in the geometric-optics limit for light waves, fermion and 
scalar wave equations.  These results motivate further investigation 
of quantum Yang-Mills gravity, its Feynman rules and the associated 
conservation of energy-momentum tensor.   

Yang-Mills gravity with Abelian group T(4) can be formulated in inertial and non-inertial frames.\cite{1,4} The basic idea is a union of space-time symmetry and gauge symmetry in the following sense:  For a general space-time symmetry of a physical system, let us consider the infinitesimal coordinate transformations,
$$x^{\mu} \to x'^{\mu}= x^{\mu}+\Ld^{\mu}(x),$$
where $\Ld^{\mu}(x)$ is an arbitrary infinitesimal vector function. The conventional view is to regard it as the mathematical coordinate transformations in Riemann geometry and to require the general covariance of the laws of nature, following the early ground-breaking work of Einstein and Grossmann.\cite{5,6}  As a result, the physics of gravity is closely related to the Riemann-Christoffel curvature tensor of the physical space-time.  In contrast, we take a new view which is more in harmony with the Yang-Mills approach.  We regard it as the localization of the constant four-dimensional translation in flat space-time, $x^{\mu} \to x'^{\mu}= x^{\mu}+\Ld_o^{\mu}$, i.e., replacing the constant vector $\Ld_o^{\mu}$ by an arbitrary infinitesimal vector function $\Ld^{\mu}(x)$.  This view leads to the Yang-Mills gravity with the local translational gauge symmetry based on the  flat space-time with arbitrary coordinates (for general non-inertial frames).\cite{1,4}  The physics of Yang-Mills gravity turns out to be closely related to the gauge curvature $C^{\mu\nu\a}$ of the translational group rather than the space-time curvature, where the gauge curvature involves a tensor gauge field $\phi_{\mu\nu}$.

The formulation of Yang-Mills gravity follows closely to that 
of quantum electrodynamics with Abelian group U(1) in flat space-time.   In 
contrast to other gauge groups for gravity,\cite{7,8,9,10} this external translation 
symmetry group T(4) in flat space-time leads to the following 
important properties of Yang-Mills gravity:

(A)  In Yang-Mills gravity, the gauge potential is a tensor field $\phi_{\mu\nu}$ in 
flat space-time, in contrast to the usual vector gauge potentials in 
Yang-Mills theory.  This difference is due to the property that the T(4) generators are $T_{\mu}=\p/\p x^{\mu}$, which do not  have the constant matrix representations.  The T(4) gauge covariant derivative is $\Delta_{\mu} = \p_{\mu} +g\phi_{\mu}^{\nu}T_{\nu}=J_{\mu}^{\nu}\p_{\nu}$, where $J_{\mu}^{\nu} =\delta_{\mu}^{\nu} + g\phi_{\mu}^{\nu}$, in inertial frames.  The gauge curvature is obtained from $[\Delta^{\mu},\Delta^{\nu}]=C^{\mu\nu\a}\p_\a$, where $C^{\mu\nu\a}= J^{\mu\s}(\p_{\s} J^{\nu\a}) -J^{\nu\s}( \p_{\s} J^{\mu\a}).
$  In this sense, Yang-Mills gravity is a generalization of the usual Yang-Mills theory from internal groups to the external space-time translational group.

(B) Yang-Mills gravity in flat space-time possesses the well defined and conserved 
energy-momentum tensor which is related to the source of the gravitational field.

(C)  Yang-Mills gravity has only attractive force between all 
matter-matter, matter-antimatter and antimatter-antimatter, because 
the T(4) group generators are $T_{\mu}=\p/\p x^{\mu}$.   

(D)  The gravitational coupling constant $g$ for a 
fermion-fermion-graviton vertex must have the dimension of 
`length', in sharp contrast to the dimensionless coupling constant in
the usual Yang-Mills theory with 
internal gauge group.  This is due to the fact that the generators 
$\p/\p x^{\mu}$ of 
the translational group has the dimension of (1/length).  Yang-Mills 
gravity suggests 
that the gravitational constant with the dimension of length is as 
fundamental as the dimensionless coupling constants in gauge field 
theories. 

(E)  Yang-Mills gravity suggests that apparently curved space-time shown by the observed motions of classical objects is a manifestation of translational gauge symmetry.  
  The reason is that an effective Riemann metric tensor emerges in the Hamilton-Jacobi 
equation if and only if one takes the classical limit (i.e., the limit of geometric optics)
of  fermions and photons wave equations with translational gauge symmetry.  
 
\section{\large Gauge invariant action for fermion and gravitational 
fields}
\noindent

  Let us consider 
quantum Yang-Mills gravity in inertial frames (with the Minkowski 
metric tensor $\eta^{\mu\nu}=(1,-1,-1,-1)$) for simplicity.  The 
gauge invariant action $S_{\phi\psi}$ for spin 2 field $\phi_{\mu\nu}=\phi_{\nu\mu}$ 
and fermions is given by\cite{1}
\be
S_{\phi\psi}=\int (L_{\phi} + L_{\psi} + 
 L_{\xi\z}) d^{4}x, \ \ \ \   c= \hbar = 1,
\ee
\be
L_{\phi}= \frac{1}{2g^2}\left (\frac{1}{2}C_{\mu\nu\a}C^{\mu\nu\a}- 
C_{\mu\a}^{ \ \ \  \a}C^{\mu\b}_{ \ \ \  \b} \right), 
\ee
\be
L_{\psi}=\frac{i}{2}[\overline{\psi} \g_\mu (\Delta^\mu 
\psi) - 
(\Delta^\mu \overline{\psi}) \g_\mu \psi] -
m\overline{\psi}\psi,
\ee
\be
L_{\xi\z}=\frac{\xi}{2g^{2}}\left[\p^\mu J_{\mu\a} - 
\frac{\z}{2} \p_\a J^\ld_\ld\right]\left[\p_\nu J^{\nu\a} - 
\frac{\z}{2} \p^\a J^\ld_\ld\right].
\ee
\be
\Delta_\mu \psi = J_{\mu\nu}\p^\nu \psi, \ \ \ \ \ \  
J_{\mu\nu}=\eta_{\mu\nu}+ 
g \phi_{\mu\nu} = J_{\nu\mu} , 
\ee
where the T(4)  gauge curvature $C^{\mu\a\b}$ is given by
$$
C^{\mu\nu\a}= J^{\mu\s}\p_{\s} J^{\nu\a}-J^{\nu\s} \p_{\s} J^{\mu\a}.
$$
The gauge-fixing term $L_{\xi\z}$ is necessary for quantization of 
fields with gauge symmetry and it
may involve two arbitrary gauge parameters 
$\xi$ and $\z$ in general.
Here, we have chosen the usual gauge condition of the form
\be
\p^\mu J_{\mu\nu} - \frac{\zeta}{2} \p_\nu J  = Y^{\nu}, \ \ \ J=
J^{\ld}_\ld=\delta^{\mu}_{\mu}- g\phi, \ \ \  \phi=\phi^{\ld}_\ld,
\ee
where $Y^{\nu}$ is a suitable function of spacetime.

The Lagrangian for pure gravity $L_{pg}=L_{\phi}+
L_{\xi\z}$ in terms of the tensor field $\phi_{\mu\nu}$ can be written as:
\be
L_{pg}= L_{2}+L_{3} +L_{4} +L_{\xi\z},
\ee
where
\be
L_{2}=\frac{1}{2}\left(\p_{\ld} \phi_{\a\b} \p^{\ld} \phi^{\a\b}\right.-
\p_{\ld} \phi_{\a\b} \p^{\a} \phi^{\ld\b}-
\p_{\ld} \phi\p^{\ld} \phi \ \ \ \ \ \ \ \ 
\ \ \ \ \
\ee
$$+2\p_{\ld} \phi \p^{\b} \phi^{\ld}_{\b}- 
\p_{\ld} \phi^{\ld}_{\mu} \p_{\b} \phi^{\mu\b}),$$
\be
L_{3}=g[(\p_{\ld} \phi_{\a\b} \p_{\rho} 
\phi^{\a\b})\phi^{\ld\rho}
-(\p_{\ld} \phi_{\a\b} \p_{\rh} 
\phi^{\ld\b})\phi^{\a\rho}-(\p_{\ld} \phi \p_{\rho} 
\phi)\phi^{\ld\rho} 
\ee
$$
+(\p^{\ld} \phi \p_{\rho} 
\phi_{\ld\b})\phi^{\b\rho} 
+(\p_{\ld} \phi \p^{\b} 
\phi^{\mu}_{\b})\phi^{\ld}_{\mu}
-(\p^{\ld} \phi_{\ld\mu} \p_{\rh} 
\phi^{\mu}_{\b})\phi^{\b\rho} ],$$
\be
L_{4}=\frac{g^{2}}{2}[(\p^{\ld} \phi_{\a\b} \p_{\rho} 
\phi^{\a\b})\phi_{\ld\mu}\phi^{\mu\rh}
-(\p_{\ld} \phi_{\a\b} \p_{\rho} 
\phi^{\mu\b})\phi^{\ld}_{\mu}\phi^{\a\rho} \ \ \ \ \ \ \ \ \ \ \ \ \
\ee
$$
+(\p_{\ld} \phi \p_{\rho} 
\phi^{\mu}_{\b})\phi^{\ld}_{\mu}\phi^{\b\rho}+(\p_{\ld} \phi \p_{\rho} 
\phi^{\mu}_{\b})\phi^{\rho}_{\mu}\phi^{\b\ld}$$
$$
-(\p_{\ld} \phi \p_{\rho} 
\phi)\phi^{\ld}_{\mu}\phi^{\mu\rho}
-(\p_{\ld} \phi_{\a\mu} \p_{\rho} 
\phi^{\mu}_{\b})\phi^{\ld\a}\phi^{\b\rho}],
$$
\be
L_{\xi\z} =\frac{\xi}{2}\left[ (\p_{\ld}\phi^{\ld\a})\p^{\rh}\phi_{\rh\a} -\zeta (\p_{\ld}\phi^{\ld\a})\p_{\a}\phi + \frac{\zeta^{2}}{4}(\p^{\a} \phi)\p_{\a} \phi \right].   
\ee
The action (1) with the 
Lagrangians given in (2), (3) and (4) leads to 
the tensor field equation in inertial 
frames,\cite{1}
\be
H^{\mu\nu} + \xi A^{\mu\nu}= g^2 T^{\mu\nu},
\ee
$$H^{\mu\nu} \equiv   \ Sym \ [\p_\ld (J^{\ld}_\rho C^{\rho\mu\nu} - J^\ld_\a 
C^{\a\b}_{ \ \ \ \b}\eta^{\mu\nu} + C^{\mu\b}_{ \ \ \ \b} J^{\nu\ld})    $$
$$- C^{\mu\a\b}\p^\nu J_{\a\b} + C^{\mu\b}_{ \ \ \ \b} \p^\nu J^\a_\a -
 C^{\ld \b}_{ \ \ \ \b}\p^\nu J^\mu _\ld],$$
$$A^{\mu\nu}=  \ Sym \ [\p^{\mu}\p_{\s} J^{\s\nu} - 
\frac{\z}{2}\p^{\mu}\p^{\nu}J - 
\frac{\z}{2}\e^{\mu\nu}\p^{\ld}\p^{\s}J_{\ld\s} + 
\frac{\z^{2}}{4}\e^{\mu\nu}\p^{2}J],  $$ 
where "Sym" denotes that $\mu$ and $\nu$ should be made symmetric.  We have the 
identities for the gauge curvature,
\be
C_{\mu\nu\a}= -C_{\nu\mu\a}, \  \  \  \  \  \   C_{\mu\nu\a} +  
C_{\nu\a\mu} +  C_{\a\mu\nu} = 0.
\ee
The symmetric `energy-momentum tensor' $T^{\mu\nu}$ in equation (12) is given by
\be
T^{\mu\nu} = \frac{1}{4}\left[ \overline{\psi} i\gamma^\mu \p^\nu \psi -
i(\p^\nu \overline{\psi}) \gamma^\mu \psi 
+ \overline{\psi} i\gamma^\nu \p^\mu \psi -
i(\p^\mu \overline{\psi}) \gamma^\nu \psi \right].  
\ee
For weak fields in inertial frames without having the gauge-fixing 
terms (i.e., setting $\xi=0$ in eq. (12)), the 
field equation can be linearized as follows:
\be
\p_\ld \p^\ld \phi^{\mu\nu} - \p^\mu \p_\ld \phi^{\ld\nu} - 
\e^{\mu\nu}\p^{2}\phi + \e^{\mu\nu}\p_{\a}\p_{\b}\phi^{\a\b} 
\ee
$$ + \ \p^\mu \p^\nu \phi - \p^\nu \p_\ld \phi^{\ld\mu} = g T^{\mu\nu}, $$
where $g$ is related to Newtonian constant G, $g=\sqrt{8\pi G}$.
In such a weak field limit,  the linearlized eq. (15) is the same as 
that in Einstein's gravity.  

In the presence of spacetime gauge field, the Dirac equation of a 
fermion can be derived from the fermion Lagrangian (3),
\be
i(\g^{\nu}+ g\g_{\mu}\phi^{\mu\nu})\p_{\nu}\psi 
+\frac{ig}{2}(\p_{\nu}\phi^{\mu\nu})\g_{\mu}\psi - m\psi=0,
\ee
 \be
i(\p^{\mu}+ g \phi^{\mu\nu} \p_{\nu})\overline{\psi}\g_{\mu} 
+\frac{ig}{2}\overline{\psi}\g_{\mu}(\p_{\nu}\phi^{\mu\nu})    
+ m\overline{\psi}=0,
\ee
It follows from these two equations that the fermion current
$j^{\nu}$ in Yang-Mills gravity is conserved,
\be
\p_{\nu}j^{\nu}= 0,
\ee
where
$$j^{\nu} =\overline{\psi} \g^{\nu}\psi 
+ g\phi^{\mu\nu}\overline{\psi} \g^{\nu}\psi
=J^{\mu\nu}\overline{\psi} \g_{\mu}\psi.$$

The conserved total energy-momentum tensor $T^{\mu\nu}(\phi\psi)$ of the graviton-fermion 
system is given by the Lagrangian $L_{\phi\psi}=L_{\phi} + L_{\psi}$,
\be
T^{\mu\nu}(\phi\psi)= T(\phi)^{\mu\nu} + T(\psi)^{\mu\nu},
\ee
$$T(\phi)^{\mu\nu} = \frac{\p L_{\phi}}{\p(\p_{\mu}\phi_{\a\b})}
\p^{\nu}\phi_{\a\b} -\e^{\mu\nu}L_{\phi}$$
$$T(\psi)^{\mu\nu} = \frac{\p L_{\psi}}{\p(\p_{\mu}\psi)}\p^{\nu}\psi
+ \p^{\nu}\overline{\psi}\frac{\p L_{\psi}}{\p(\p_{\mu}
\overline{\psi})} -\e^{\mu\nu}L_{\psi}, $$
according to Noether's theorem.  Note that $L_{\psi}$ does not contain $\p_{\mu}\phi_{\a\b}$ and $L_{\phi}$ does not involve $\p_{\mu} \psi$ and $\p_{\mu} \overline{\psi}$.  We find that
\be
T(\phi)^{\mu\nu} = \frac{1}{g}(J_{\ld}^{\mu}C^{\ld\a\b} - 
\e^{\a\b}J_{\ld}^{\mu}C^{\ld\s}_{ \ \ \ \s} +
J^{\b\mu}C^{\a\s}_{ \ \ \ \b})\p^{\nu} 
\phi_{\a\b}
\ee
$$- \e^{\mu\nu}\frac{1}{4g^{2}}(C_{\mu\a\b}C^{\mu\a\b}- 
2C_{\mu\a}^{ \ \ \  \a}C^{\mu\b}_{ \ \ \  \b}),$$
\be
T(\psi)^{\mu\nu}= \frac{i}{2}[ \overline{\psi} \gamma_{\a}J^{\a\mu} \p^\nu \psi -
(\p^\nu \overline{\psi}) J^{\a\mu} \gamma_\a  \psi] - \e^{\mu\nu} L_{\psi}.
\ee
These conserved energy-momentum tensors of the fermion and gauge fields are not symmetric, but they
can be symmetrized, if one wishes. \cite{11,12,13}   

We note that the non-symmetric  
energy-momentum tensor $T(\psi)^{\mu\nu}$ of fermions differs from the symmetric
energy-momentum tensor $T^{\mu\nu}$ in (12) and (14), which acts as the 
source for producing the 
gravitational field $\phi_{\a\b}$.  
From the linearized equation (15), one can derived the conservation law
\be 
 \p_{\mu}T^{\mu\nu}=0.
\ee
It is gratifying that the source tensor $T^{\mu\nu}$  in the field equation (12) does not involve the tensor gauge field, in contrast to the Noether tensor $T^{\mu\nu}(\psi)$ in (21).  Otherwise, 
the fermion-graviton coupling and the high-energy behavior of amplitudes will be more complicated.  

\section{The Feynman rules for quantum Yang-Mills gravity}
\noindent

Let us consider the general propagator of the graviton corresponds to the 
gauge-fixing terms in (4) that involve two arbitrary parameters $\xi$ and 
$\zeta$.  It can be derived from the free Lagrangians (8) and (11) which 
contains $\xi$ and $\zeta$.  We find that
\be
G_{\a\b,\rh\s}=-i\left(\frac{1}{k^{2}}
\left[\e_{\a\b}\e_{\rh\s}- \e_{\rh\a}\e_{\s\b}-\e_{\rh\b}\e_{\s\a}\right]
-\frac{1}{k^{4}}\frac{2\zeta-2}{\zeta-2}\right.
\ee
$$\times \left(k_{\a}k_{\b}\e_{\rh\s}\right.
+\left.k_{\rh}k_{\s}\e_{\a\b}\right)
+\frac{1}{k^{4}}\frac{\xi+2}{\xi}(k_{\s}k_{\b}\e_{\rh\a}
+k_{\a}k_{\s}\e_{\rh\b}+ k_{\rh}k_{\b}\e_{\s\a}$$
$$\left.+k_{\rh}k_{\a}\e_{\s\b})
-\frac{1}{k^{6}}\frac{(2\zeta-2)}{\xi(\zeta-2)}
\left[\frac{4-2\xi+4\xi\zeta}{\zeta-2}-4-2\xi\right]
k_{\a}k_{\b}k_{\rh}k_{\s}\right),$$
where the usual prescription of $i\ep$ for the Feynman propagator is 
understood.
In a special case for $\zeta=1$ and arbitrary $\xi$, the propagator 
(23) reduced to a simpler form
\be
G_{\a\b,\rh\s}=-i\left[\frac{1}{k^{2}}(\e_{\a\b}\e_{\rh\s}-
\e_{\rh\a}\e_{\s\b}-\e_{\rh\b}\e_{\s\a})\right.
\ee
$$
+ \left.\frac{1}{k^{4}}\frac{\xi+2}{\xi}(k_{\s}k_{\b}\e_{\rh\a}
+k_{\a}k_{\s}\e_{\rh\b}+ k_{\rh}k_{\b}\e_{\s\a}+
k_{\rh}k_{\a}\e_{\s\b})\right].$$
For $\zeta=0$ and arbitrary $\xi$, the graviton propagator (23) 
reduces to the 
form
\be
G_{\a\b,\rh\s}=-i\left(\frac{1}{k^{2}}[\e_{\a\b}\e_{\rh\s}-
\e_{\rh\a}\e_{\s\b}-\e_{\rh\b}\e_{\s\a}] \right.
\ee
$$
- \frac{1}{k^{4}}[k_{\a}k_{\b}\e_{\rh\s}
+k_{\rh}k_{\s}\e_{\a\b}]
+ \frac{1}{k^{4}}\left[\frac{\xi+2}{\xi}\right](k_{\s}k_{\b}\e_{\rh\a}
+k_{\a}k_{\s}\e_{\rh\b}$$
$$+ k_{\rh}k_{\b}\e_{\s\a}
+k_{\rh}k_{\a}\e_{\s\b})
+ \left.\frac{1}{k^{6}}
\left[\frac{\xi+6}{\xi}\right]
k_{\a}k_{\b}k_{\rh}k_{\s}\right).$$
These special cases are consistent with that obtained in  previous 
works of Fradkin and Tyutin,\footnote{ Their free Lagrangian with linearized form of harmonic gauge condition differs from the present free Lagrangian by a factor of (1/2), so that their gauge parameter $\a$ in the graviton propagator is related to $\xi $ in equations (4) and (24) by the relation $1/\a =\xi/2$.
}\cite{14} and others.\footnote{The graviton propagator obtained in refs. 15 and 16 corresponds to the special case $\xi=-2$ in (24).}\cite{15,16,17}

Let us consider the Feynman rules for graviton interaction corresponding  to
$iL_{3}$.  The momenta in the Feynman rules in this paper are in-coming to the vertices.
Suppose the tensor-indices and momenta of the 3-vertex are
denoted by $[\phi^{\mu\nu}(p)\phi^{\s\tau}(q)\phi^{\ld\rh}(k)]$.  We obtain the graviton 3-vertex,
\be
ig \ Sym \ P_{6}\left(\frac{}{}-p^{\ld}q^{\rho}\e^{\s\mu}\e^{\tau\nu}
+p^{\s}q^{\rho}\e^{\ld\mu}\e^{\tau\nu}+p^{\ld}q^{\rh}\e^{\s\tau}\e^{\mu\nu}\right.
\ee
$$ 
\left.-p^{\s}q^{\rho}\e^{\mu\nu}\e^{\tau\ld} 
 -p^{\rho}q^{\s}\e^{\mu\nu}\e^{\tau\ld}
+p^{\mu}q^{\rho}\e^{\nu\s}\e^{\ld\tau}\frac{}{}\right).$$
We use the symbol Sym to denote a symmetrization is to be performed on each index 
pair $(\mu\nu)$, $(\s\tau)$ and $(\ld\rh)$  in $[\phi^{\mu\nu}(p)\phi^{\s\tau}(q)\phi^{\ld\rh}(k)]$.
The symbol  $P_{n}$ denotes a summation is 
to be carried out over permutations of the momentum-index 
triplets, and the subscript gives the number of permutations 
in each case.  For example, we have
$$
  P_6 (p^{\ld} q^{\rho}\eta^{\s\mu}\eta^{\tau\nu})=   (p^{\ld} q^{\rho}\eta^{\s\mu}\eta^{\tau\nu}+p^{\s} k^{\tau}\eta^{\ld\mu}\eta^{\rho\nu}
$$
$$+p^{\rho} q^{\ld}\eta^{\mu\s}\eta^{\nu\tau}+p^{\tau} k^{\s}\eta^{\mu\ld}\eta^{\nu\rho}+q^{\mu} k^{\nu}\eta^{\ld\s}\eta^{\rho\tau}+k^{\mu} q^{\nu}\eta^{\s\ld}\eta^{\tau\rho}), $$
which is the result that one obtains directly from the first term in the Lagrangian $L_3$ in (9) by following the usual method for obtaining the vertex in Feynman rules.\cite{18,19}

The 4-vertex (i.e., $iL_{4}$ with 
$\phi^{\mu\nu}(p)\phi^{\s\tau}(q)\phi^{\ld\rh}(k) \phi^{\a\b}(l)$) is given by
\be
\frac{ig^{2}}{2}Sym  \ P_{24}\left(\frac{}{}-p^{\rh} 
q^{\b}\e^{\mu\s}\e^{\nu\tau}\e^{\ld\a}+ 
p^{\rh}q^{\b}\e^{\ld\s}\e^{\nu\tau}\e^{\mu\a}\right.
\ee
$$
- p^{\rh}q^{\b}\e^{\ld\s}\e^{\tau\a}\e^{\mu\nu}
- p^{\b}q^{\rh}\e^{\ld\s}\e^{\tau\a}\e^{\mu\nu}
$$
$$\left.+p^{\rh}q^{\b}\e^{\mu\nu}
\e^{\s\tau}\e^{\ld\a}
+p^{\rh}q^{\b}\e^{\ld\nu}\e^{\tau\a}\e^{\mu\s}\frac{}{}\right).$$

The fermion propagator has the usual form
\be
\frac{i}{\g^{\mu} p_{\mu} - m}.
\ee
The fermion-graviton 3-vertex (i.e., $iL_{\psi}^{int}$ with 
$\overline{\psi}(q) \psi(p) \phi_{\mu\nu}(k)$) takes the form
\be
Sym \ \frac{i}{2} g\g_{\mu}(p_{\nu} + q_{\nu}),
\ee
where $\psi(p)$ may be considered as an annihilation operator of a 
fermion with the momentum $p_{\nu}$ and $\overline{\psi}(q)$ a creation 
operator of a fermion with the momentum $q_{\nu}$.

The ghost fields and their 
interactions with the graviton  
in Yang-Mills gravity can be obtained by the Faddeev-Popov 
method.\cite{20,21} 
Let us choose the gauge condition (6) for discussions.  It is 
important to take care of symmetrization of indices related to the 
indices of the symmetric tensor field.  Thus we write the gauge 
condition (6) in the following form where the indices $\mu$ and $\nu$ of 
$J_{\mu\nu}$ are explicitly symmetrized: 
\be
Y_{\ld}=\frac{1}{2}(\d^{\mu}_{\ld}\p^{\nu} + \d^{\nu}_{\ld}\p^{\mu} - 
\z \e^{\mu\nu}\p_{\ld})J_{\mu\nu}
\ee
where $Y^{\nu}$ is independent of 
the fields and the gauge function $\L^{\mu}$.  The vacuum-to-vacuum 
amplitude of quantum Yang-Mills gravity is
\be
W(Y^{\ld})= \int d[\phi_{\a\b}, \overline{\psi}, \psi]
\left(exp\left[i\int d^{4}x(L_{\phi}+ L_{\psi})\right] \right.
\ee
$$\left. \times (det \ U) \mbox{\large \boldmath $ \Pi$}_{x,\nu}\d(\p_{\mu}J^{\mu\nu}-\frac{\z}{2}\p^{\nu}J-Y^{\nu})\right) .$$
This is similar to that in the usual Yang-Mills theory.\cite{22}

The functional determinant $det \ U$ is determined by
\be
\frac{1}{det \ U}= \int d[\L^{\ld}]\mbox{\large \boldmath $ \Pi$}_{x,\nu}
\de\left(\frac{1}{2}(\d^{\mu}_{\ld}\p^{\nu} + \d^{\nu}_{\ld}\p^{\mu} - \z \e^{\mu\nu}\p_{\ld})J^{\$}_{\mu\nu}-Y_{\ld}\right),
\ee
where $J^{\$}_{\mu\nu}$ is given by the T(4) gauge 
transformation\cite{1}
\be
J^{\$}_{\mu\nu} = J_{\mu\nu} -\L^{\ld} \p_{\ld} J_{\mu\nu} 
-(\p_{\mu}\L^{\ld})J_{\ld\nu}-(\p_{\nu}\L^{\ld})J_{\mu\ld}.
\ee
It follows from (32) and (33) that
\be
U_{\mu\nu}=
(\p^{\ld} E_{\mu\nu\ld}) + E_{\mu\nu\ld} \p^{\ld},
\ee
where
$$
E_{\mu\nu\ld}=J_{\mu\nu}\p_{\ld}+ J_{\nu\ld}\p_{\mu}+ 
(\p_{\nu}J_{\mu\ld}) - \z \e_{\mu\ld}J_{\nu\s}\p^{\s}
-\frac{\z}{2}\e_{\mu\ld} (\p_{\nu}J).$$

The vacuum-to-vacuum amplitude can be written in the form
\be
W = \int d[Y^{\nu}(x)] W(Y^{\ld})
exp\left[-i\int d^{4}x\left( \frac{\xi}{2g^{2}}Y^{\mu}Y_{\mu}\right)\right],
\ee
$$=\int d[\phi_{\a\b}, \overline{\psi}, \psi]detU 
exp\left[i\int d^{4}x \{L_{\phi}+L_{\psi} + L_{\xi\z}\right],$$
to within an unimportant multiplicative factor.

As usual, the functional determinant $det U$ in (32) can be expressed in terms of an 
effective Lagrangian for fictitious
vector-fermion fields $V^{\mu}$ and $\overline{V}^{\mu}$,
\be
det(U_{\mu\nu})=\int exp \left(i\int L_{eff} d^{4}x\right) 
d[V(x)^{\ld},\overline{V}(x)^{\s}],
\ee
$$L_{eff} = \overline{V}^{\mu} U_{\mu\nu}
V^{\nu},$$
where $U_{\mu\nu}$ is given by (34) and $\overline{V}^{\nu}$ 
is considered as an independent field.
The quanta of the vector-fermion fields $V^{\mu}$ and $\overline{V}^{\nu}$ in the 
effective Lagrangian $L_{eff}$ are the (Faddeev-Popov) ghost particles and anti-ghost 
particles in Yang-Mills gravity.  The effective Lagrangian in (36) is 
consistent with those obtained by the Lagrangian multiplier 
method.\cite{22} 

Thus, effectively the Yang-Mills gravity of fermions and gravitons is 
given by the following total Lagrangian $L_{tot}$,
\be
L_{tot}=L_{\phi}+L_{\psi}+L_{\xi\z}+\overline{V}^{\mu}U_{\mu\nu}
V^{\nu}.
\ee
To obtain the Feynman rules for the general
propagator (involving an arbitrary gauge parameter $\zeta$) for the ghost 
particle and the interaction vertices
of the ghost 
particle, it is convenient to write the effective Lagrangian in
(36) in the following form
\be
L_{eff}= \overline{V}^{\mu}\left[\e_{\mu\nu}\p_{\ld}\p^{\ld}+(1-\zeta)\p_{\mu}\p_{\nu}\right]V^{\nu}
\ee
$$
- g\left[(\p^{\a}\overline{V}^{\b}) V^{\nu}\p_{\nu}\phi_{\a\b} +   (\p^{\ld}\overline{V}^{\mu})(\p_{\ld} V^{\nu})\phi_{\mu\nu}
+(\p^{\ld}\overline{V}^{\mu})(\p_{\mu} V^{\nu})\phi_{\ld\nu}\right] 
$$
$$
+g \zeta(\p^{\ld}\overline{V}_{\ld})(\p^{\mu} V^{\nu})\phi_{\mu\nu} + g(\zeta/2) (\p_{\mu}\overline{V}^{\mu}) V^{\nu} (\p_{\nu}\phi_{\a\b})\e^{\a\b}.
\ $$
 
The propagator of the ghost vector particle can be derived from the 
effective Lagrangian in (36) for ghost field,
\be
G^{\mu\nu}=\frac{-i}{k^{2}}
\left(\e^{\mu\nu}-\frac{k^{\mu} k^{\nu}}{k^{2}}\frac{(1-\zeta)}{(2-\zeta)}\right)
\ee
Similarly, the ghost-ghost-graviton vertex (denoted by 
$\overline{V}^{\mu}(p)V^{\nu}
(q)\phi^{\a\b}(k)$) is found to be
\be
\frac{ig}{2}[p^{\a}k^{\nu}\eta^{\mu\b} + p^{\b}k^{\nu}\eta^{\mu\a} +
p\cdot q(\eta^{\mu\a}\eta^{\nu\b}+ 
\eta^{\mu\b}\eta^{\nu\a})+p^{\a}q^{\mu}\eta^{\nu\b}
\ee
$$ 
+ p^{\b}q^{\mu}\eta^{\nu\a}-\zeta p^{\mu}k^{\nu}\eta^{\a\b}
- \zeta p^{\mu}q^{\b}\eta^{\nu\a} - \zeta 
p^{\mu}q^{\a}\eta^{\nu\b}].$$
We have used the convention that all momenta are incoming to the 
vertices.  All ghost-particle vertices are 
bilinear in the vector-fermions, as shown in the effective Lagrangian 
(36).  Thus the vector-fermion appears, 
by definition of the physical subspace, only in closed loops in the 
intermediate states of a physical process, and 
there is a factor of -1 for each vector-fermion loop in Yang-Mills 
gravity.

\section{Effective curved space-time for motions of classical objects 
and effective Riemann metric tensor in Hamilton-Jacobi equation}

For a logically consistent theory of Yang-Mills gravity, the equation 
of motion for classical objects or matter must follow from the wave 
equation (16) for fermions---the fundamental constituents of matter.
This connection is necessary from the physical viewpoint.  It is 
gratifying that Yang-Mills gravity allows the spacetime gauge field 
$\phi_{\mu\nu}$ of quantum particles in a classical object to add up 
coherently when one considers the classical limit (or the 
geometric-optics limit) of the fermion wave equation (16) in flat spacetime.

In order to see this connection between quantum and classical 
equations, we write the fermion field $\psi$ in the usual eikonal form
\be
\psi = \psi_{o} exp[iS].
\ee
We have the following simple property
\be
\psi_{t}= \mbox{\large \boldmath $ \Pi$}_{k} (\psi_{ok} exp[iS_{k}])
=\psi_{to}exp[iS_{t}],
\ee
$$\psi_{to}= \mbox{\large \boldmath $ \Pi$}_{k}\psi_{ok}, \ \ \ \  
S_{t}=\sum_{i=1}^{N} S_{k}$$
Let us write the fermion in the presence of gravitational field 
$\phi_{\mu\nu}$ in the following form
\be
i\g_{\mu}\left[\e^{\mu\nu}+g\phi^{\mu\nu}\p_{\nu}+\frac{1}{2}\p_{\nu}J^{\mu\nu}\right]\psi - m\psi=0,
\ee
\be
(\g_{\mu}\g_{\nu}+\g_{\nu}\g_{\mu}) =2\e_{\mu\nu}, \ \ \ \ 
\e_{\mu\nu}=(+1,-1,-1,-1)
\ee
It follows from (41) and (43) that
\be
\g_{\mu}(\e^{\mu\nu} + g\phi^{\mu\nu})\p_{\nu}S - 
\frac{ig}{2}\g_{\mu}\p_{\nu}\phi^{\mu\nu} + m =0.
\ee
In the classical limit, the mass m, the eikonal S and its spacetime 
derivatives (or the 4-momentum) $\p_{\mu}S$ are respectively 
very much larger than the gravitational field, the Planck constant 
$\hbar$ and the spacetime derivatives of $\phi_{\mu\nu}$,
\be
m, \ \  S, \ \ \  \p_{\mu}S >> \phi^{\mu\nu}, \ \ \  \hbar, \ \ \  g\p_{\nu}\phi^{\mu\nu}
\ee
where $m$ and $\p_{\mu}S$ have the dimension of (1/length) 
in the natural units $c=\hbar=1$.
Thus, we have the equation
\be
\g_{\mu}(\e^{\mu\nu} + g\phi^{\mu\nu})\p_{\nu}S + m = 0
\ee
in the classical limit.  Following Dirac, in order to eliminate the 
spin effect associated with $\g_{\mu}$, one may multiply a factor
$[\g_{\mu}(\e^{\mu\nu} + g\phi^{\mu\nu})\p_{\nu}S - m] $ to (47) from 
left.  We obtain the classical equation
\be
G^{\mu\nu}(\p_{\mu}S)(\p_{\nu}S) - m^{2} = 0
\ee
\be
G^{\mu\nu} \equiv (\e^{\mu\s} + g\phi^{\mu\s})(\delta^{\nu}_{\s} + 
g\phi^{\nu}_{\s}).
\ee
This is precisely the relativistic Hamiltonian-Jacobi equation of 
motion for classical objects in Yang-Mills gravity.  The presence of the `effective Riemann 
metric tensor' $G^{\mu\nu}$ in the Hamilton-Jacobi equation appears to
indicate that a classical object moves in a curved spacetime.  
However, the real physical spacetime for fields and quantum particles is 
flat.  From the viewpoint of Yang-Mills theory, such an apparent 
`curved spacetime' is nothing but the classical manifestation of the 
fermion-graviton interaction with the 
spacetime translation gauge symmetry.

From the steps (44)-(46), one sees formally the result 
and implications of large mass of a 
classical object.  However, the large mass and large eikonal S in the 
classical limit, as shown in (47) are physically due to the summation 
of $m_{k}$ and $S_{k}$ for each quantum particle k in the system 
under consideration.  They are not coming from the 
mathematical process of  taking the 
limits $m \to \infty$ and $\hbar \to 0$.  (Note that in the 
non-natural 
units, the phase $iS_{k}$ in (42) should be replaced by 
$iS_{k}/\hbar$).  In this sense,
the conventional steps (44)-(46) do not directly reveal the physical  connection 
between a macroscopic body with a large mass $m$ and 
that with  a large  number of constituent 
particles' mass $m_{k}$, k=1,2,3,...N for a N-particle system, namely
\be
m_{t} = \sum_{k=1}^{N} m_{k}.
\ee
Following the same steps from (41) to (47), we have
 the relation for the particle $k$,
\be
[\g_{\mu}(\e^{\mu\nu} + g\phi^{\mu\nu})\p_{\nu}S]_{(k)} + 
m_{(k)} = 0, \ \ \ \   k=1,2,...N.
\ee
After, one sums over all particles, one obtains the expression 
(47) with m replaced by $m_{t}$ in (50), 
\be
\g_{\mu}\{\e^{\mu\nu} + g\phi^{\mu\nu}(x)\}\p_{\nu}S(x) + m_{t} = 0.
\ee
We also make natural requirements 
for classical limit: 
Namely, the momentum of particles in a macroscopic object add up 
`coherently' in the following sense,
\be
\sum_{k=1}^{N}[\p_{\nu}S]_{(k)}=\p_{\nu}S(x), \ \ \ \
\sum_{k=1}^{N}[\phi^{\mu\nu}\p_{\nu}S]_{(k)}=\phi^{\mu\nu}(x)\p_{\nu}S(x).
\ee
The results in (52) and (53) appear to be natural because the classical 
macroscopic object is, by construction, made 
of the N bounded particles.   
Although the summation of constant masses in (50) is simple, the 
summation of,say, spacetime-dependent momenta in the first equation 
of (53) within the 
relativistic framework is 
not, because each particle has its own time and space coordinates in 
a general inertial frame.  Perhaps, a simple picture to visualize 
this equation in (53) is that 
the relative motions of quantum particles can be ignored in the classical 
limit.  They can be viewed as at rest relative to each other 
and, hence, can be effectively described by using 
only one spacetime coordinates.   From 
this viewpoint, the passage from a system of bounded quantum 
particles to one single macroscopic classical object is not as simple as one might 
think.

Once one has the result (47) or (52), one can ignore the spin effect 
by removing $\g_{\mu}$ from the equation, as done in (47)-(48).
In this way, one obtains the 
relativistic Hamilton-Jacobi equation (48) with an effective Riemannian 
metric tensor $G^{\mu\nu}$, whose presence in the classical limit
is crucial for the Yang-Mills gravity to be 
consistent with experiments.

\section{Discussions and remarks}

At a firs glimpse, the Lagrangian for the electromagnetic field in 
the presence of gravitational field $\phi_{\mu\nu}$ appears to be 
quite different from the Lagrangian (3) of the Dirac field.  However, if 
one follows the Yang-Mills approach with T(4) gauge group, i.e., 
replacing $\p^{\mu}$ by T(4) gauge covariant derivative
$\p^{\mu}+g\phi^{\mu\nu}\p_{\nu}$, one will also obtain a 
relativistic Hamilton-Jacobi equation (48) with $m=0$ for light rays 
in the limit of geometric optics.\cite{1}

Tensor fields with spin 2 was discussed by Fierz and Pauli in 
1939.\cite{23,24}  In order to have 2 independent polarization 
states for a massless symmetric tensor field `$\psi_{\mu\nu}$,' certain 
subsidiary conditions were imposed.  In particular, they discussed 
tensor field equations which are invariant under a 
`gauge transformation', 
$$\psi_{\mu\nu} \to \psi_{\mu\nu} + 
\p_{\mu}\L_{\nu} + \p_{\nu}\L_{\mu}.$$
Without the guidance of modern 
gauge symmetry, it is non-trivial to find a Lagrangian for tensor 
fields in such a way that wave equations and subsidiary conditions 
follow simultaneously from the Hamilton principle.\cite{11}  It turns 
out that their simplest free Lagrangian is equivalent to $L_{2}$ in (8), 
and their free field equation for massless $\psi_{\mu\nu}$
is exactly the same as linearized equation (15) in Yang-Mills gravity 
and in Einstein's theory of gravity.

It is generally believed that general coordinate invariance in 
Einstein's gravity can be considered as a gauge symmetry.  In 
particular, one can use the affine connection to construct covariant 
derivatives of tensors, just as one uses gauge field to construct 
gauge covariant derivatives of matter fields.  However, a fundamental 
difference between Einstein's gravity and Yang-Mills gravity is as 
follows:  Einstein's gravity is assumed to be based on curved 
spacetime, so that the 
commutator of two covariant derivatives with respect to $x^{\mu}$ 
and $x^{\nu}$ leads to the Riemann-Christoffel curvature tensor 
uniquely and unambiguously.  On the contrary, Yang-Mills gravity is 
assumed to be based on T(4) group in flat spacetime, just 
like that in the usual Yang-Mills theory and conventional field 
theories.
This implies that, in Yang-Mills gravity, there is 
no Riemann-Christoffel curvature and that there is only T(4) gauge 
curvature $C^{\mu\nu\ld}$.

The fact that the action in
Yang-Mills theory involves quadratic gauge curvature, while
Hilbert-Einstein action involves linear curvature of spacetime, implies a basic 
and important break down of their analogy.\cite{25}  Namely, gauge fields are 
fundamental and are not expressed in terms of any more fundamental 
fields, while the affine connection in Einstein's gravity is itself 
constructed from first derivatives of the metric 
tensor.  In this connection, we stress that there is no 
break down of analogy  between Yang-Mills gravity and Yang-Mills 
theory.

There is a formulation of gravity called teleparallel gravity (TG),\cite{26,27} which is based on the translational gauge symmetry on a flat space-time with a torsion tensor.  In TG, a flat connection with torsion makes translations local gauge symmetries and whose curvature tensor is the torsion field.  However, the concept of translational gauge symmetry is realized differently in Yang-Mills gravity and in TG.  Although there is a one-one correspondence between equations in Yang-Mills gravity and teleparallel gravity, there is a true difference between them.  Their differences show up in the properties of the gauge potentials, gauge curvature and actions in inertial frames.

\noindent
(a) Gauge covariant derivative. 
 
In Yang-Mills gravity, the $T(4)$ gauge covariant derivative  
$ \D_{\mu}\psi $  is defined in (5), where the gauge potential field $\phi_{\mu\nu}$ and $J_{\mu\nu}$ are symmetric tensors in flat space-time.  In contrast, the gauge covariant derivative in teleparallel gravity is defined by\cite{18}
\be
D_{\mu}\psi=\p_{\mu}\psi +B_{\mu}^{a}\p_{a} \psi = h^{a}_{\mu}\p_{a} \psi,  \ \ \ \ \   h^{a}_{\mu} = \p_{\mu} x^a + B^a_{\mu}.
\ee
 where $x^{\mu}$ is the coordinates of space-time with the metric tensor $g_{\mu\nu}(x)=\e_{ab}h^a_{\mu}(x) h^b_{\nu}(x)$, and  $x^{a}(x^{\mu})$ is the coordinates in a tangent Minkowski space-time with the metric tensor $\e_{ab}=(+,-,-,-).$  Thus, $h^{a}_{\mu}$ in TG is treated as a tetrad rather than a tensor.

\noindent 
(b) Gauge curvature.

In Yang-Mills gravity, the gauge curvature $C^{\mu\nu\a}$ is a tensor of the third rank and is defined by
\be
[\D_{\mu}, \D_{\nu}]= C_{\mu\nu \ld}\p^{\ld}, \ \ \ \
C^{\mu\nu\ld}= J^{\mu\s}\p_{\s} J^{\nu\ld}-J^{\nu\s} \p_{\s} J^{\mu\ld}.
\ee
$$J_{\mu\nu}=\eta_{\mu\nu}+ 
g \phi_{\mu\nu} = J_{\nu\mu}.$$
On the other hand, the gauge curvature in TG is a torsion tensor $T^a_{\mu\nu}$:
\be
  [D_{\mu}, D_{\nu}]= T^a_{\mu\nu}\p_{\a}, \ \ \ \
T^a_{\mu\nu}=\p_{\mu} h^a_{\nu}-  \p_{\nu} h^a_{\mu}.
\ee

\noindent
(c)  Gravitational action.

The real physical difference between Yang-Mills gravity and teleparallel gravity shows up in their gravitational actions.  The gravitational Lagrangian $L_{\phi}$ in Yang-Mills gravity is given in equation (2).  In TG,  the gravitational action and Lagrangian $L_{h}$ are given by Hayashi and Shirafuji\cite{26,27}
\be
S_{TG}=\int  L_{h}\sqrt{-det  \ g_{\mu\nu} } \ d^{4}x,
\ee
\be
L_{h}= a_1 (t_{\ld\mu\nu}t^{\ld\mu\nu}) + a_2 (v^{\mu} v_{\mu}) + a_3 (a^{\mu}a_{\mu}),
\ee
$$
 t_{\ld\mu\nu}=\frac{1}{2}(T_{\ld\mu\nu} +T_{\mu\ld\nu}) +\frac{1}{6}(g_{\nu\ld}v_{\mu} + g_{\nu\mu}v_{\ld}) -\frac{1}{3} g_{\ld\mu} v_{\nu} $$
 $$
v_{\mu} =T^{\ld}_{ \ \  \ld\mu}, \ \ \ \ \    a_{\mu} = \frac{1}{6} \epsilon_{\mu\nu\rho\s} T^{\nu\rho\s}. $$

$$T^{\ld}_{ \ \   \mu\nu}=h^{\ld}_{a}(\p_{\nu} h^a_{\mu} - \p_{\mu} h^a _{\nu}), \ \ \ \ \ \ \   g_{\mu\nu}(x)=\e_{ab}h^a_{\mu}(x) h^b_{\nu}(x)
$$
where $T^{\ld}_{ \ \   \mu\nu}$ is the  torsion  tensor.  It appears impossible to adjust the three parameters $a_1, \ a_2$ and $a_3$ in (58) to make $S_{TG}$ in (57) to be the same as $S=\int L_{\phi} d^4x$ in Yang-Mills gravity for both inertial and non-inertial frames.\cite{4, 1}   In general, if the tetrad in a field theory involves dynamical fields, it will lead to  complicated vertices in Feynman rules due to the presence of $\sqrt{-det  \ g_{\mu\nu}}$  in the action, if the field can be quantized.

Yang-Mills gravity is a local gauge field theory for microscopic world and, hence, is not related to the equivalence principle in general.  However, in the classical limit (i.e., geometric-optics limit), the wave equation of fermion reduces to a relativistic Hamiltonian-Jacobi equation (48)  which can describe the motion of a free-fall classical object in inertial frames.  In this sense, Yang-Mills gravity is compatible with the equivalence principle in the geometric-optics limit.

Recently Ning Wu proposed to interpret the Lagrangian in 
Hilbert-Einstein action for gravity in terms of a gauge 
field with a T(4) gauge symmetry in flat 
space-time.\cite{28}  In this way, the function $g_{\mu\nu}$ (or the 
Riemannian space-time) is 
originated from the presence of a physical tensor 
gauge-field.  A similar idea was also discussed by Logunov and his 
collaborators.\cite{29,30}  
But in their formulations of gravity, the graviton still 
involves N-vertex of self-coupling, where N is an arbitrarily
large number and, hence, is  more complicated 
than the maximum 4-vertex  for self-coupling of graviton in the present Yang-Mills gravity.  Because the simplicity of 
graviton's self-couplings  in Yang-Mills gravity, we expect 
this theory to have a better 
high-energy behavior for the S-matrix.\cite{31,32,33}  These properties merit 
further study.

\bigskip

\noindent
{\bf Acknowledgements}
\bigskip

The work was supported in part by the Jing Shin
Research Fund of the UMass Dartmouth Foundation.  The author would like to thank Leonardo Hsu for his collaboration at earlier stage.  He is a  substantial contributor to the space-time symmetry (in inertial and non-inertial frames) for $T(4)$ gauge symmetry and  Yang-Mills gravity.

\bigskip

\bigskip

\bigskip

\bigskip

\noindent
{\large \bf Appendix} 
 
\bigskip 
\noindent  
{\bf Yang-Mills gravity and its consistence with experiments}
\bigskip

The present quantum Yang-Mills gravity is based on a simple but non-trivial union of space-time translational symmetry and local gauge symmetry.  The theory has a new T(4)  translational gauge-invariant action $S_{\phi\psi}$ [ see  eq. (A1) below] in flat space-time with arbitrary coordinates.  Thus, it differs from all previous non-metric Yang-Mills type formulations of gravity proposed long time ago and ruled out by different reasons.\cite{7,8,9,10}   The T(4) translational gauge-invariant action $S_{\phi\psi}$ assures that the wave equations of the fermion and the electromagnetic  fields reduce to a classical Hamilton-Jacobi equation with an  `effective metric tensor' [see eq. (A2) below] in the geometric-optics limit (i.e., the classical limit).  This is very interesting because even though the underlying physical space-time for quantum particles and fields is flat,  the classical objects moves as if it is in a curved space-time with the `effective metric tensor' (which is determined by the physical field $\phi_{\mu\nu}$).  

Thus, Yang-Mills gravity suggests a new understanding of gravity:  Gravity can be understood on the basis of the T(4) `gauge curvature' $C_{\mu\nu\a}$ in equation (2), and its Lagrangian involves a quadratic gauge curvature.  Furthermore, gravity can be understood within the framework of flat space-time.  This property is very important  because it enables one to carry out the usual quantization procedure for gravitational field and it allows a genuine conservation of energy-momentum tensor.   In contrast, the conventional theory is based on curved space-time, which appears to be too general to have a well-defined conservation law for the energy-momentum tensor and to carry out a satisfactory  quantization of the gravitational field.

The gauge-invariant non-linear field equation (12) with $\xi=0$ for $\phi_{\mu\nu}$ is highly symmetric, so that it is a nuisance to solve explicitly  even a static solution order by order.  One has to take into account of the gauge-fixing terms $ \xi A^{\mu\nu}$ in (12) and show that the observable results are independent of the gauge parameter $\xi$.\cite{1}  To avoid misunderstanding, let us summarize the novel features of Yang-Mills gravity and its consistence with experiments as follows:\cite{1,2}
	
Yang-Mills gravity is based on the mathematical framework of flat space-time with arbitrary coordinates and a metric tensor $P_{\mu\nu}$.  The gravitational action (1) is written in inertial frames for simplicity of quantization and Feynman rules.  However, the classical theory can be formulated in inertial and non-inertial frames (i.e., general frames with a metric tensor $P_{\mu\nu}$\cite{4}) with the gauge invariant action $S_{\phi\psi}$ involving fermions and gravitational fields\cite{1} 
$$
\hspace{1in}S_{\phi\psi}= \int (L_{\phi}+L_{\psi}) \sqrt{-P} d^4x, \ \ \ \ \    P=det \ P_{\mu\nu}, \ \ \ \ \ \ \ \ \ \ \ \ \ \ \ \ (A1)
$$
where $L_{\phi}$ and $L_{\psi}$ are given in equations (2) and (3), and
$$C^{\mu\nu\a}= J^{\mu\s}D_{\s} J^{\nu\a}-J^{\nu\s} D_{\s} J^{\mu\a}.$$
$$
\Delta^\mu \psi = J^{\mu\nu}D_\nu \psi, \ \ \ \ \  
J^{\mu\nu}=P^{\mu\nu}+
g \phi^{\mu\nu} = J^{\nu\mu} , \ \ \ \ \
D_\ld P_{\mu\nu}=0.
$$ 
In a general frame of reference, the ordinary partial derivatives in the gauge curvature $C_{\mu\nu\a}$ in (2) and the fermion Lagrangian (3) are replaced by the covariant derivative $D_\mu$ 
 associated with the metric tensor $P_{\mu\nu}$.   One can understand $P_{\mu\nu}$ as the general metric tensor for the space-time of non-inertial frames and it reduces to the Minkowski metric tensor $\eta_{\mu\nu}$ in the limit of zero acceleration.\cite{1}  We note that the gauge-fixing term
$L_{\xi\z}$ must remain the same form as (4) [involving ordinary partial differentiations] in a general coordinate in order to fix the chosen gauge.

In the geometric-optics limit, the wave equations for a spinor or a vector (or electromagnetic) field reduces to the Hamilton-Jacob equation with the form\cite{1,2}
$$
\hspace{0.5in}G^{\mu\nu}(\p_\mu S)(\p_\nu S) - m^2 = 0, \ \ \ \ \ \ \   G^{\mu\nu}=P_{\a\b}J^{\a\mu}J^{\b\nu},  \ \ \ \ \ \ \   m \ge 0. \ \ \   (A2)
$$
The 'effective' metric tensor $G^{\mu\nu}$ emerges if and only if one takes the classical limit for the motion of macroscopic objects.  It is not an inherent property of the physical space-time in general because it is determined by the physical tensor field $\phi_{\mu\nu}$ through the relation $J^{\mu\nu}=P^{\mu\nu}+g \phi^{\mu\nu} $ for $G^{\mu\nu}$ in (A2).  

Yang-Mills gravity predicts that, in inertial frames, the perihelion shift has a new small correction term in the second order:
$$
\hspace{0.7in} \delta \phi = \frac{6\pi Gm}{P} \left(1 - 
\frac{3(E_o^2 - m_p^2)}{4 m_p^2} \right), \ \ \ \ \ \ \  P= \frac{M^2}{m_p^2 Gm},    \ \ \ \ \ \ \ \ \ \  (A3)
$$
where $M$ and $m_p$ are respectively the angular momentum and the mass of the planet.\cite{1} 
The second term in the bracket of (A3) shows the difference between  the present Yang-Mills gravity and Einstein's theory.  This small correction term cannot be detected in the solar system with the present observational accuracy.  In the future, the result (A3) may be possible to test outside the solar system by, say, the observation of  the quasar OJ 287.  This quasar contains a binary black hole system where one black hole pierces the accretion
disk of the second and produces periodic outbursts.  The system has v/c = 0.1 and could be used to test deviations from Einstein's theory of gravity.\cite{34,35}

Moreover, the bending of light has also a new correction term in the second order:\cite{1}
$$
\hspace{0.7in}\Delta \phi =  \frac{4Gm\omega_o}{M'} \left(1 - 
\frac{18 G^{2}m^{2} 
\omega_o^{2}}{M'^{2}} \right). \ \ \ \ \ \ \ \  M'=\omega_o R  \ \ \ \ \ \ \ \ \  (A4)
$$
The additional correction term in the bracket differs from 
that in Einstein's gravity and is too small to be detected for the bending of visible light by the Sun. 

Following the usual method  and 
approximation, one can calculate the power radiated by a body rotating
around one of 
the principal axes of the ellipsoid of inertia.\cite{36}  At twice the
 rotating frequency $\Omega$,
 i.e., $\omega=2\Omega$, we obtain the total power $P_o(\omega)$
emitted by a rotating body:\cite{2}
$$
 \ \ \ \ \ \ \ \ \ \ \ \ \ \ \ \ \  P_o(2\Omega) = \left[ \frac{32G_{N} \Omega^6 I^2 e_{q}^2}{5} 
\right], \ \ \ \ \ \   c=1. \ \ \ \ \ \ \ \ \ \ \ \ \ \ \ \ \ \ \ \ \ \ \ \ \ \ \  (A5)
$$
where $I$ and $e_{q}$ are respectively moment of inertia and equatorial 
ellipticity.  Thus,  the
 gravitational quadrupole radiation (A5) 
obtained in Yang-Mills gravity is  the same as 
that of Einstein's gravity to the second order approximation.\cite{2}


\bibliographystyle{unsrt}

\end{document}